\documentclass[11pt]{article}

\usepackage{amsmath}
\usepackage{amssymb}
\usepackage{graphicx}
\usepackage{cite}
\input{epsf}
\newcommand{\beq}{\begin{equation}}
\newcommand{\eeq}{\end{equation}}
\newcommand{\beqa}{\begin{eqnarray}}
\newcommand{\eeqa}{\end{eqnarray}}

\begin{document}

%%\noindent Accepted for publication in Phys. Lett. {\bf B} 

\begin{flushright}
{\tiny{FZJ-IKP-TH-2009-15}}
{\tiny{HISKP-TH-09/17}}
\end{flushright}

\vspace{.6in}

\begin{center}

\bigskip

{{\Large\bf A method to measure the antikaon-nucleon scattering
    length\\[0.3em] 
    in lattice QCD
}\footnote{
Work supported in part by DFG (SFB/TR 16,
``Subnuclear Structure of Matter'') and  by the Helmholtz Association
through funds provided to the virtual institute ``Spin and strong
QCD'' (VH-VI-231). We also acknowledge the support of the European
Community-Research Infrastructure Integrating Activity ``Study of
Strongly Interacting Matter'' (acronym HadronPhysics2, Grant
Agreement n. 227431) under the Seventh Framework Programme of EU.
A.R. acknowledges financial support 
of the Georgia National Science Foundation (Grant \#GNSF/ST08/4-401).}}

\end{center}

\vspace{.3in}

\begin{center}
{\large 
Michael Lage$^\ddagger$\footnote{email: lage@itkp.uni-bonn.de},
Ulf-G. Mei{\ss}ner$^\ddagger$$^\ast$\footnote{email: meissner@itkp.uni-bonn.de},
Akaki Rusetsky$^\ddagger$\footnote{email: rusetsky@itkp.uni-bonn.de}
}

\vspace{1cm}

$^\ddagger${\it Universit\"at Bonn,
Helmholtz--Institut f\"ur Strahlen-- und Kernphysik (Theorie) and\\
Bethe Center for Theoretical Physics, Universit\"at Bonn,
D-53115 Bonn, Germany}

\bigskip

$^\ast${\it Forschungszentrum J\"ulich, Institut f\"ur Kernphysik 
(Theorie), J\"ulich Center for Hadron Physics\\
and Institute for Advanced Simulation, D-52425 J\"ulich, Germany}

\bigskip

\bigskip

\end{center}

\vspace{.4in}

\thispagestyle{empty} 

\begin{abstract}\noindent 
We propose a method to determine the isoscalar  $\bar K N$ scattering length 
on the lattice. Our method represents the generalization of L\"uscher's 
approach in the presence of inelastic channels (complex scattering length).
In addition, the proposed approach allows one to find the position of the 
S-matrix pole corresponding the the $\Lambda(1405)$ resonance.  
\end{abstract}

\vfill

\pagebreak

%%%%%%%%%%%%%%%%%%%%%%%%%%%%%%%%%%%%%%%%%%%%%%%%%%%%%%%%%%%%%%%%%%%%%%%
\noindent {\bf 1.} 
The antikaon-nucleon scattering amplitude is of fundamental importance in 
nuclear, particle and astrophysics, see e.g. 
Refs.~\cite{Friedman:2007zz,Weise:2007rd,Brown:2007ara,Magas:2008bp}.
In particular, the $\bar KN$ system at threshold provides an interesting 
testing ground of the chiral dynamics of QCD with strange quarks due 
to the $\Lambda (1405)$ resonance just below the scattering threshold.
In fact, experimental information  on the $K^- p$ scattering length 
from scattering data and kaonic hydrogen level shifts is contradictory, 
as first stressed in \cite{Meissner:2004jr} and further elaborated on in
Refs.~\cite{Borasoy:2004kk,Oller:2006jw,Borasoy:2006sr}. A clarification is
expected from the upcomimg SIDDHARTA experiment at DA$\Phi$NE, that 
intends to remeasure kaonic hydrogen with unprecedented accuracy and
is expected to give further constraints to the isoscalar and isovector
kaon-nucleon scattering lengths from the first measurement of the
energy spectrum of kaonic deuterium. 

\medskip\noindent
On the theoretical side, effective field theory methods in various disguises
are employed to pin down the $\bar K N$ scattering length. The scattering
information is usually analyzed in terms of unitarized versions of chiral
perturbation theory, that  lead to a dynamic generation of various resonances,
in particular the $\Lambda (1405)$. Such schemes have been worked out over
the years by various groups at various levels of sophistication 
(see, e.g.~\cite{Borasoy:2004kk,Oller:2006jw,Borasoy:2006sr,KN}). 
While an impressive amount of data (cross sections, threshold ratios, mass
distributions, etc.)  is described in such  approaches with good precisison, 
unitarization of course introduces some unwanted model-dependence. 
On the other hand, the  extraction of the $\bar K N$ 
scattering length from kaonic hydrogen is firmly rooted in non-relativistic
bound-state effective field theory and thus is devoid of the 
above-mentioned model-dependence (for a recent review, see
\cite{Gasser:2007zt}). However,  
it entirely rests on the availablity of precise
kaonic atom data. Unfortunately,  the existing data from DEAR \cite{Beer:2005qi} 
and KEK \cite{Ito:1998yi} are conflicting. It would therefore be most
welcome to have another tool at hand that would allow one to determine this
fundamental quantity. 

\medskip\noindent
As we will argue in this letter, lattice QCD provides
such a framework. As first shown by L\"uscher, finite volume simualations
of the energy levels of two-particle states can give access to scattering 
information \cite{Luscher:1986pf,Luscher:1990ux}.
The idea of L\"uscher is very elegant and simple. For two-particle states that 
are well separated from bound states or resonances in the given channel, the
$1/L$ expansion of the energy levels takes the generic form (here, $L$ is the
size of the box with volume $L\times L\times L$)
\beq
E \sim \frac{a}{L^3} \left\{ 1 + c_1 \frac{a}{L} + c_2 \frac{a^2}{L^2}\right\}
+ {\cal O}(L^{-6})~, 
\eeq
where $a$ is the desired scattering length and $c_1, c_2$ are pure
numbers (see below).  The method  has e.g. been used to
extract the $\pi\pi$, $\pi K$ and $KN$ $S$-wave scattering lengths
from lattice data \cite{Beane:2007xs}.  Note that an alternative proposal
to extract the scattering length from the two-particle wave function is
e.g. given in Ref.~\cite{Aoki:2005uf}.

\medskip\noindent
However, for the extraction
of the $\bar KN$ scattering length, a generalization of this scheme is
called for since there is a strong channel coupling between $\bar KN$ 
and $\Sigma \pi$, the latter channel having its threshold about 100~MeV 
below the opening of the $\bar K N$ one. In addition, the appearance of
the $\Lambda (1405)$ just between these two thresholds further complicates
the picture. All these features can be captured by a two-channel 
Lippmann-Schwinger
equation. As we will show in the following, a suitable formulation of this
equation in the finite volume allows for an unambigous extraction of the
complex-valued isoscalar $K^- N$ scattering length. Note also that our method
is similar to the approach adopted in Ref.~\cite{He}, where the problem was 
treated within the potential scattering theory. In this paper, we use instead 
the language of the non-relativistic effective field theory (NR EFT), which
enables one to systematically address the effects of particle 
creation/annihillation and relativistic corrections.

\medskip

\begin{figure}[t]
\centering
\includegraphics[width=0.3\textwidth]{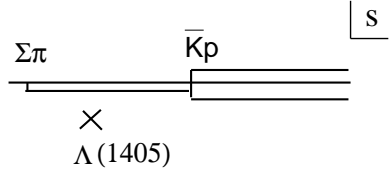}
\caption{
Complex $s$-plane with the $\Sigma \pi$ and $\bar KN$ cuts along the
real axis and the location of the $\Lambda(1405)$ resonance.
}
\label{fig:anstruc}
\end{figure}

%%%%%%%%%%%%%%%%%%%%%%%%%%%%%%%%%%%%%%%%%%%%%%%%%%%%%%%%%%%%%%%%%%%%%%%%%%%%%%
\noindent {\bf 2.}
To set the stage, we consider a two-channel Lippmann-Schwinger (LS) 
equation 
in NR EFT in the
infinite volume. Note that we are using a covariant version of the 
NR EFT, considered in Refs.~\cite{cuspwe}.
The channel number 1 refers to $\bar KN$ and 2 to
$\Sigma\pi$ with total isospin $I = 0$.
The resonance $\Lambda (1405)$ is located between
two thresholds, on the second Riemann sheet, close to the real axis
(these thresholds are defined by 
$s_t = (m_N + M_K)^2$ and 
$s'_t=(m_\Sigma+M_\pi)^2$)~\footnote{In the following, we work in
  the  isospin limit and thus do not resolve the further splitting of these
  thresholds.}.
\noindent
Consider first energies above $\bar KN$ 
threshold, $s > (m_N + M_K)^2$. The
coupled-channel LS equation for the $T$-matrix elements
$T_{ij}(s)$ in the dimensionally regularized 
NR EFT reads (we only consider $S$-waves here)
\beqa\label{eq:LSinfini}
T_{11}  &=& H_{11} + H_{11} \, iq_1 T_{11} + H_{12} \, iq_2
T_{21}~,\nonumber\\
T_{21}  &=& H_{21} + H_{21} \, iq_1 T_{11} + H_{22} \, iq_2
T_{21}~,
\eeqa
with $q_1 = \lambda^{1/2} (s,m_N^2,M_K^2)/(2\sqrt{s})$, 
$q_2 = \lambda^{1/2} (s,m_\Sigma^2,M_\pi^2)/(2\sqrt{s})$ and 
$\lambda(x,y,z)$ stands for the K\"allen function.
Furthermore, the $H_{ij}(s)$ denote the driving potential 
in the corresponding channel.
Continuation of the center-of-mass momentum 
$q_1$ below threshold $(m_\Sigma + M_\pi)^2 <s< (m_N + M_K)^2$
is obtained via (see Fig.~\ref{fig:anstruc} for the corresponding analytical
structure)
\beq
iq_1 \to -\kappa_1 = -\frac{(-\lambda(s,M_K^2,m_N^2))^{1/2}}{2\sqrt{s}}
\eeq
The resonance corresponds to a pole on the second Riemann sheet in the
complex $s$-plane, its
position can de determined from the secular  equation
\beq\label{eq:secular}
\Delta (s) = 1 + \kappa_1^R \, H_{11} - \kappa_2^R \, H_{22} - 
\kappa_1^R \kappa_2^R \, \left( H_{11} H_{22} - H_{12}^2\right)
\eeq
with $\kappa_1^R = -(-\lambda (s_R,m_N^2,M_K^2))^{1/2}/(2\sqrt{s_R})$ and 
$\kappa_2^R = (-\lambda (s_R,m_\Sigma^2,M_\pi^2)^{1/2})/(2\sqrt{s_R})$.
The energy and width of the resonance are then  given by $\sqrt{s_R} = E_R - i \Gamma_R/2$.
The $\bar K N$ scattering length is related to the amplitude $T_{11}$ at $s = s_t
= (m_N+M_\pi)^2$ via
\beq\label{scattlength}
a_{11}  \equiv T_{11} (s_t) = H_{11} (s_t) + \frac{iq_2 (s_t) \,
  (H_{12} (s_t))^2}{ 1 - i q_2 (s_t) \, H_{22} (s_t) }~.
\eeq
Thus, to pin down its complex value, we need to determine the three real 
quantities
$H_{11}, H_{12}, H_{22} $ at $s=s_t$ appearing  in Eq.~(\ref{scattlength})

\medskip

%%%%%%%%%%%%%%%%%%%%%%%%%%%%%%%%%%%%%%%%%%%%%%%%%%%%%%%%%%%%%%%%%%%%%%%%%%%%%%%%%%%%%%%%
\noindent {\bf 3.} 
We now consider the same problem in a finite volume. The rotational symmetry
is broken to a cubic symmetry so that the infinite volume version of the LS
equation Eq.~(\ref{eq:LSinfini}) takes the form (we consider only
$S$-waves here, neglecting the small mixing to higher partial waves. The mixing
can be easily included at latter stage, see e.g., Ref.~\cite{Lage}.),
\beqa
T_{11}  &=& H_{11} - \frac{2}{\sqrt{\pi}L}\, Z_{00}(1;k_1^2)\,  H_{11}
T_{11} - \frac{2}{\sqrt{\pi}L}\, Z_{00}(1;k_2^2)\,  H_{12} T_{21}~,\nonumber\\
T_{21}  &=& H_{21} - \frac{2}{\sqrt{\pi}L}\, Z_{00}(1;k_1^2)\,  H_{21}
T_{11} - \frac{2}{\sqrt{\pi}L}\, Z_{00}(1;k_2^2)\,  H_{22} T_{21}~,\nonumber\\
\eeqa
with 
\beqa
k_1^2 &=& \left(\frac{L}{2\pi}\right)^2 \, \frac{\lambda(s,M_K^2,m_N^2)}{4s}~, \nonumber\\
k_2^2 &=& \left(\frac{L}{2\pi}\right)^2 \, \frac{\lambda(s,M_\pi^2,m_\Sigma^2)}{4s}~, \nonumber\\
Z_{00} (1;k^2) &=& \frac{1}{\sqrt{4\pi}} \,\lim_{r\to 1}\sum_{\vec{n} \in R^3}
\frac{1}{({\vec n\,}^2 - k^2)^r}~.
\eeqa
Here, we have neglected the terms that vanish exponentially at a large 
$L$.
%(a regularization is implicit in the expression of $Z_{00} (1;k^2)$, see below).
%Now if $q^2 = -b^2, b^2 > 0$, then
%\beq
%Z_{00} (1; -b^2) = -\pi^{3/2} \, b\, \left\{ 1 - \frac{1}{2\pi}
%  \sum_{|\vec{j}| \neq 0} \frac{1}{|\vec{j}|\, b} {\rm
%    e}^{-2\pi\,|\vec{j}|\,b} \right\}~. 
%\eeq
%\medskip
The secular equation that determines the spectrum can be brought into the
form
\beqa\label{eq:pseudophase}
&&\qquad  1 - \frac{2}{\sqrt{\pi} L} \,  Z_{00}(1;k_2^2)\, F(s,L) = 0~, \nonumber\\
&& F(s,L) = \left[ H_{22} -   \frac{2}{\sqrt{\pi}L}\, Z_{00}(1;k_1^2)\,  
(H_{11}H_{22} - H_{12}^2)\right] \, \left[1 -  \frac{2}{\sqrt{\pi}L}\, 
Z_{00}(1;k_1^2)\,  H_{11}\right]^{-1}
\eeqa
This is rewritten as 
\beqa
\delta(s,L) &=& -\phi(k_2) + n\,\pi~, \quad n = 0,1,2, \ldots  \nonumber\\
\phi(k_2) &=& -\arctan \frac{\pi^{3/2} \, k_2}{ Z_{00} (1;k_2^2)}~,
\eeqa
with
\beq
\tan\delta(s,L) = q_2(s) \,  F(s, L)~.  
\eeq
$\delta (s,L)$ is called the {\em pseudophase}.
It is a function of the energy $\sqrt{s}$ and the level index 
$n$, $\delta_n (s) = \delta(s,L_n(s))$. 

\medskip\noindent
The dependence of the pseudophase on $s$ and $L$ (or, equivalently, on 
the level index $n$) is very different from that of the usual scattering 
phase. Namely, the elastic phase extracted from the lattice data by using
L\"uscher's formula is independent of the volume modulo terms that 
exponentially vanish at a large $L$. Further, the energies where the
phase passes through $\pi/2$ lie close to the real resonance locations.
In contrast with this, the pseudophase contains  terms which are only
power suppressed at a large $L$. Moreover, it contains the tower of 
resonances which are not related to the dynamics of the system
in the infinite volume and merely reflect the existence of the 
discrete energy levels in the ``shielded'' channel.

\begin{figure}[t]
\centering
\includegraphics[width=0.68\textwidth]{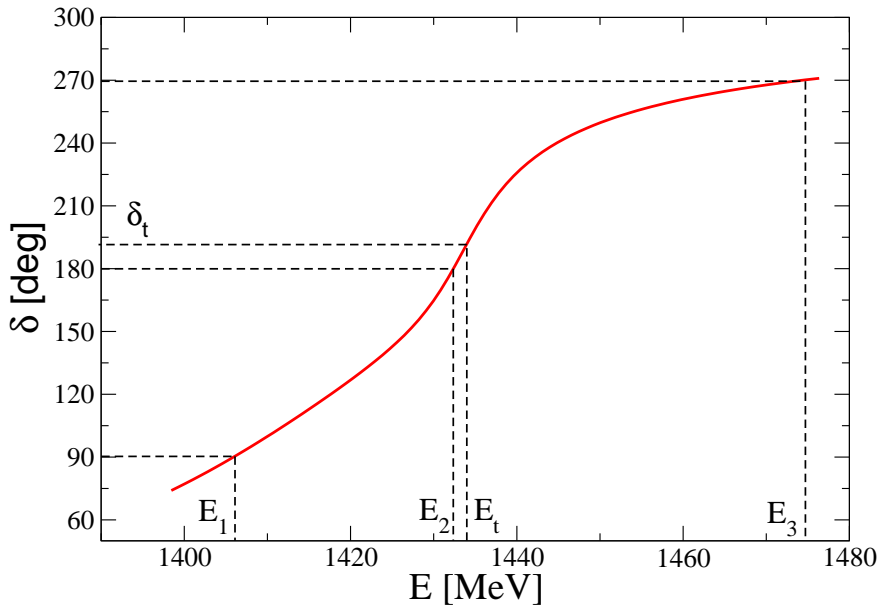}
\vspace{2mm}
\caption{
The  pseudophase $\delta$.  
The energy $E_1$ at which the pseudophase passes through $\pi/2$
corresponds to the $\Lambda(1405)$.
}
\label{fig:phase}
\end{figure}

\begin{figure}[t]
\centering
\includegraphics[width=0.64\textwidth]{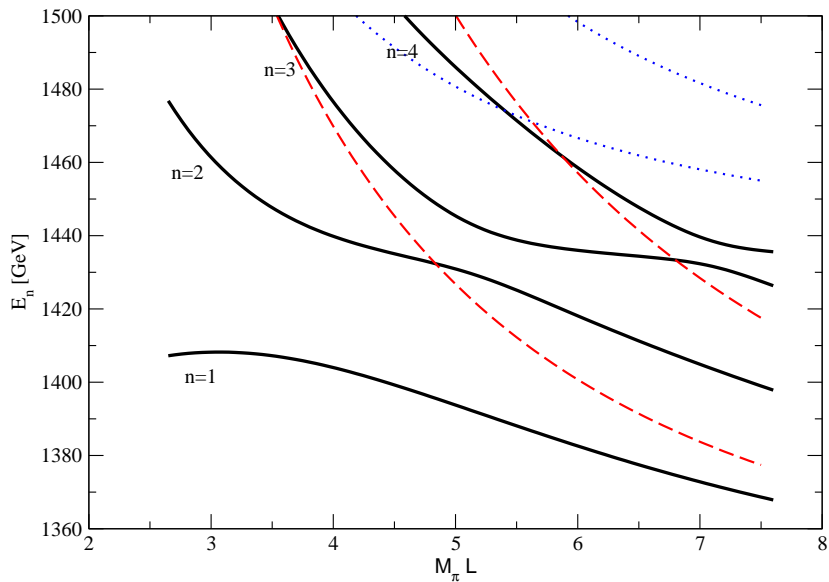}
\caption{
Energy levels for the two-channel model with an explicit $\Lambda (1405)$
resonance in the finite volume.  The avoided level crossing which
is observed at the energies between $1430~\mbox{MeV}$ and 
$1440~\mbox{MeV}$ is not related to the physical resonance in the 
infinite volume but reflects the presence of the $\bar KN$ threshold.
For comparison, we plot the energy levels levels for the non-interacting 
two-particle systems $\pi\Sigma$ (dashed
lines) and $\bar KN$ (dotted lines).
}
\label{fig:levels}
\end{figure}

\medskip\noindent
Measuring the pseudophase on the lattice can be used to determine
the $\bar KN$ scattering length. It can be directly 
seen from the expression of the pseudophase, 
which depends on real functions $H_{11},H_{12},H_{22}$. Extracting these
from the data, we then find the scattering length by using
Eq.~(\ref{scattlength}). 
Note that in the expression for the scattering length 
we need $H_{ij}(s)$ evaluated at
threshold $s=s_t$. We shall however demonstrate below that
replacing $H_{ij}(s)$ by $H_{ij}(s_t)$ in certain observables,
related to the pseudophase,
introduces very small correction, since the effective range term 
proportional to $(s-s_t)$ is suppressed by $L^{-3}$ as compared
to the leading  order result. To be specific, we consider the following
three observables:

\begin{enumerate}

\item
For some chosen value of $n$, we measure the value of the pseudophase 
$\delta(s_t;L(s_t))\doteq\delta_t$ at threshold 
$s_t = (m_N + M_K)^2$ and $E_t=\sqrt{s_t}$ 
(see Fig.~\ref{fig:levels}). On the other hand, we may express $\delta_t$
through $H_{ij}$ at $s=s_t$ in the following way.
At threshold, $Z_{00}$ is singular,
\beq
Z_{00} (1;k^2) = -\frac{1}{\sqrt{4\pi}} \, \frac{1}{k^2} + {\cal O}(1)~,
\eeq
so that
\beqa
F(s,L)|_{s \to s_t} &=& 
H_{22}(s_t) -  H_{12}^2(s_t)/
 H_{11}(s_t)
\\  
\tan \delta(s_t; L(s_t)) &=& q_2 (s_t) \,
\left( H_{22}(s_t) -  H_{12}^2(s_t)/
 H_{11}(s_t) \right) \doteq  q_2 (s_t) \, I(s_t)~.
\label{eq:threshold}
\eeqa
Thus, measuring $\delta_t$, we may extract the combination
$H_{22} - H_{12}^2/H_{11}$. 

\item
Suppose that $\tan \delta(s;L(s))$ is infinite at $s=s_3=E_3^2$
and $L=L_3=L(s_3)$ 
(see Fig.~\ref{fig:phase} for a specific representation of the
pseudophase based on a two-channel K-matrix model described below). 
This occurs at the energy where 
the denominator of Eq.~(\ref{eq:pseudophase}) vanishes
\beq\label{eq:eq}
1 - \frac{2}{\sqrt{\pi}L}\, Z_{00}(1;k_1^2(s_3))\,  H_{11}(s_3)=0\, .
\eeq
We solve this equation by expanding both $H_{11}(s)$ and 
$Z_{00}(1;k_1^2(s))$ in Taylor series in the vicinity of $s=s_t$
\beq\label{eq:H11taylor}
H_{11}(s) = H_{11}(s_t) + q_1^2(s)\,
H_{11}' (s_t) + {\cal O}(q^4) 
\eeq
and
\beq\label{eq:s1}
\frac{2}{\sqrt{\pi} L} \, Z_{00} (1;k^2)
= \frac{1}{\pi L} \frac{1}{k^2} + \frac{c_1}{L} + \frac{\pi k^2}{L} 
(c_1^2 - c_2) + {\cal O}(k^4) ~,
\eeq
with
\beq
c_1 =\frac{1}{\pi}\,\lim_{r\to 1}
\sum_{{\bf n}\neq 0}\frac{1}{({\bf n}^2)^r}
=-2.837297\ldots\, ,\quad\quad
%\biggl(\sum_{{\bf n}\neq 0}^\Lambda\frac{1}{{\bf n}^2}-4\pi\Lambda\biggr)
%=-2.837297\ldots\, ,\quad\quad
c_2 =c_1^2-\frac{1}{\pi^2}\,
\sum_{{\bf n}\neq 0}\frac{1}{{\bf n}^4}=6.375183\ldots\, .
\eeq
Substituting Eqs.~(\ref{eq:H11taylor}) and (\ref{eq:s1}) 
into Eq.~(\ref{eq:eq}), we obtain
\beq\label{eq:s1new}
q_1^2(s_3)
= -\frac{4\pi H_{11}(s_t)}{L_3^3} \left( 1 +
c_1 \frac{H_{11}(s_t)}{L_3} + c_2  \frac{H_{11}(s_t)^2}{L_3^2} 
+ {\cal O}\left(\frac{1}{L_3^3}\right) \right)\, .
\eeq
This means that measuring
the value of $s$, where the $\tan\delta(s;L(s))$
becomes infinite, we may extract
$H_{11}$ at $s=s_t$. Note that the effective-range term, which contains
$H_{11}' (s_t)$, contributes first at ${\cal O}(L^{-6})$.

\item
Similarly, suppose that $\tan\delta(s,L) =0$ at
$s=s_2=E_2^2$ and $L=L_2=L(s_2)$ (see Fig.~\ref{fig:phase}). 
Using the same technique as just described, we
obtain:
\beq\label{eq:s2}
q_1^2(s_2)
= -\frac{4\pi G(s_t)}{L_2^3} \left( 1 +
c_1 \frac{G(s_t)}{L_2} + c_2  \frac{G(s_t)^2}{L_2^2} 
+ {\cal O}\left(\frac{1}{L_2^3}\right) \right)~,
\eeq
where
\beq
 G(s_t) = H_{11}(s_t) - H_{12}^2(s_t) /
H_{22}(s_t)~.
\eeq
Therefore, measuring $s_2$, we get $H_{11} - H_{12}^2/H_{22}$.
\end{enumerate}

Using finally Eq.~(\ref{scattlength}), 
we can express the scattering length in terms of the
three quantities $H_{11}$, $I$ and $G$, all taken at $s = s_t$
\beqa\label{eq:central}
a_{11} =
H_{11}(s_t) + \frac{ iq_2(s_t) \,
I(s_t) H_{11}(s_t) \left( H_{11}(s_t)/G(s_t) -1 \right)}
{1 -  iq_2(s_t)\, I(s_t) H_{11}(s_t)/G(s_t)}~.
\eeqa
This is the central result of this letter. 

\medskip\noindent
Finally, note that in the analysis of the lattice data it may be more 
convenient to directly fit the explicit expression
of the pseudophase given in Eq.~(\ref{eq:pseudophase}) to the measured 
values on the lattice around $s=s_t$, replacing $H_{ij}(s)$ by 
$H_{ij}(s_t)$ and considering $H_{ij}(s_t) ~(ij=11,12,22)$
 as three independent fitting parameters. From the above discussion
one may expect that such a fit will lead to the precise determination
of these parameters.
The effective range terms can be neglected 
since their contribution is suppressed by three powers 
of $L$. In this case, one does not need to measure the 
pseudophase in the whole interval between $s_2$ and $s_3$. 

\medskip
%%%%%%%%%%%%%%%%%%%%%%%%%%%%%%%%%%%%%%%%%%%%%%%%%%%%%%%%%%%%%%%%%%%%%%%%%%%
\noindent {\bf 4.}
Given the parameters $H_{ij}$ determined from fitting to the pseudophase,
the position of the pole on the second Riemann sheet of the complex variable 
$s$, which corresponds to the $\Lambda(1405)$-resonance,
can be determined from the secular equation~(\ref{eq:secular}).
We expect that replacing $H_{ij}(s)$ by $H_{ij}(s_t)$ allows one to locate
the pole position at a reasonable accuracy.

\medskip
%%%%%%%%%%%%%%%%%%%%%%%%%%%%%%%%%%%%%%%%%%%%%%%%%%%%%%%%%%%%%%%%%%%%%%%%%%%
\noindent {\bf 5.}
In order to demonstrate the above-described proposal in practice, 
we have investigated a
coupled-channel model with an explicit $\Lambda (1405)$ resonance located
at $\mbox{Re}\,\sqrt{s_R} = 1406\,\mbox{MeV}$ and 
$-2\,\mbox{Im}\,\sqrt{s_R} = 50\,\mbox{MeV}$. Effective
range terms are neglected. The matrix elements $H_{ij}$ are taken 
equal to
\beq
H_{11}=-1.47573~\mbox{fm}\, ,\quad\quad
H_{12}=0.91581~\mbox{fm}\, ,\quad\quad
H_{22}=-0.34159~\mbox{fm}\, .
\eeq
This corresponds to  $a_{11} = a_0(K^- N) = (-1.26 + i\, 0.70)\,$fm. 
\noindent
The resulting first four energy levels as a function
of $M_\pi L$ are shown in  Fig.~\ref{fig:levels}. 
We notice that the lowest level ($n = 1$) does only show a moderate volume 
dependence in the interval considered, 
quite in contrast to the excited ones with $n\geq 2$. For 
$M_\pi L\simeq 2\ldots 3$ the ground state level flattens around
$E=1406~\mbox{MeV}$ that corresponds to the $\Lambda(1405)$. It is clear
that, for this reason, 
the lowest level can not be used for the extraction of the
$\bar KN$ scattering length.   
The excited levels show a more complicated behavior in this
interval of $L$. At the first glance, these levels 
exhibit the so-called avoided level crossing somewhere between
$1430~\mbox{MeV}$ and $1440~\mbox{MeV}$. In the elastic case, 
such a behavior of the energy levels signalizes the presence of a 
narrow resonance near this energy. However, this is not the case here.
The peculiar behavior of the excited energy levels is caused by the
opening of the $\bar KN$ threshold. At higher energies, the picture
repeats -- an avoided level crossing emerges, if the $\bar KN$ system
has a discrete eigenvalue at this energy in a finite volume. 
If the volume changes, the avoided level crossing moves (in difference
to the avoided level crossing corresponding to the ``true'' resonance).
For $L\to \infty$ the bifurcation lines accumulate at 
threshold $s=s_t$. In this limit,
the scattering amplitude is not analytic at $s=s_t$ (unitary cusp).

\medskip\noindent
In  Fig.~\ref{fig:phase} gives the
pseudophase derived from the second ($n = 2$) energy level. It shows the
expected behavior. First, it crosses $\pi/2$ at $\sqrt{s} = E_1$, very close
to the mass of the $\Lambda (1405)$. Then, it passes $\pi$ at $\sqrt{s}= \sqrt{s_2}=
E_2$, close to the threshold where its value is $\delta_t > \pi $.
At $E_2$, the tangent of the pseudophase vanishes at since $q_1^2 (s_2) <0$,
we can conclude that $G(s_t) > 0$, cf. Eq.~(\ref{eq:s2}). 
Finally, the value of $3\pi /2$  is reached at  $\sqrt{s}=\sqrt{s_3}= E_3$. 
Here, $q_1^2 (s_3)>0$ and consequently $H_{11} (s_t) <0$. This can be
deduced from  Eq.~(\ref{eq:s2}) after the substitutions $s_2 \to s_3$
and $L_2 \to L_3$.

\medskip

%%%%%%%%%%%%%%%%%%%%%%%%%%%%%%%%%%%%%%%%%%%%%%%%%%%%%%%%%%%%%%%%%%%%%%%%%
\noindent {\bf 6.} In this letter, we have generalized L\"uscher's
algorithm for the extraction of the scattering length from the 
finite-volume energy spectrum measured on the lattice. The modified
algorithm applies to the case when the scattering length is complex
due to the presence of the open channel(s) below threshold. 
In the case of the $\bar KN$ scattering with total isospin $I=0$,
the scattering length can be determined by measuring
 the volume dependence of the first excited level 
around the threshold energy.

\bigskip\bigskip

\noindent{\large {\bf Acknowledgments}}

\smallskip\noindent
We are grateful to S.~D\"urr, J.~Gasser, J.~Negele and F.~Niedermayer for 
useful discussions.

%%%%%%%%%%%%%%%%%% REFERENCES %%%%%%%%%%%%%%%%%%%%%%%%%%%%


\vskip 1cm
\begin{thebibliography}{99}

\frenchspacing

%\cite{Friedman:2007zz}
\bibitem{Friedman:2007zz}
  E.~Friedman and A.~Gal,
  %``In-medium nuclear interactions of low-energy hadrons,''
  Phys.\ Rept.\  {\bf 452} (2007) 89
  [arXiv:0705.3965 [nucl-th]].
  %%CITATION = PRPLC,452,89;%%

%\cite{Weise:2007rd}
\bibitem{Weise:2007rd}
  W.~Weise,
  %``Chiral SU(3) dynamics, anti-K N interactions and the quest for
  %antikaon-nuclear clusters,''
  arXiv:nucl-th/0701035.
  %%CITATION = NUCL-TH/0701035;%%

%\cite{Brown:2007ara}
\bibitem{Brown:2007ara}
  D.~B.~Kaplan and A.~E.~Nelson,
  %``Strange Goings on in Dense Nucleonic Matter,''
  Phys.\ Lett.\  B {\bf 175} (1986) 57;
  %%CITATION = PHLTA,B175,57;%%

  G.~E.~Brown, C.~H.~Lee and M.~Rho,
  %``Recent Developments on Kaon Condensation and Its Astrophysical
  %Implications,''
  Phys.\ Rept.\  {\bf 462} (2008) 1
  [arXiv:0708.3137 [hep-ph]].
  %%CITATION = PRPLC,462,1;%%

%\cite{Magas:2008bp}
\bibitem{Magas:2008bp}
  V.~K.~Magas, E.~Oset and A.~Ramos,
  %``Critical review of [K- ppn] bound states,''
  Phys.\ Rev.\  C {\bf 77} (2008) 065210
  [arXiv:0801.4504 [nucl-th]].
  %%CITATION = PHRVA,C77,065210;%%


%\cite{Meissner:2004jr}
\bibitem{Meissner:2004jr}
  U.-G.~Mei{\ss}ner, U.~Raha and A.~Rusetsky,
  %``Spectrum and decays of kaonic hydrogen,''
  Eur.\ Phys.\ J.\  C {\bf 35} (2004) 349
  [arXiv:hep-ph/0402261].
  %%CITATION = EPHJA,C35,349;%%

%\cite{Borasoy:2004kk}
\bibitem{Borasoy:2004kk}
  B.~Borasoy, R.~Nissler and W.~Weise,
  %``Kaonic hydrogen and K- p scattering,''
  Phys.\ Rev.\ Lett.\  {\bf 94} (2005) 213401
  [arXiv:hep-ph/0410305].
  %%CITATION = PRLTA,94,213401;%%

%\cite{Oller:2006jw}
\bibitem{Oller:2006jw}
  J.~A.~Oller,
  %``On the strangeness -1 S-wave meson baryon scattering,''
  Eur.\ Phys.\ J.\  A {\bf 28} (2006) 63
  [arXiv:hep-ph/0603134].
  %%CITATION = EPHJA,A28,63;%%

%\cite{Borasoy:2006sr}
\bibitem{Borasoy:2006sr}
  B.~Borasoy, U.-G.~Mei{\ss}ner and R.~Nissler,
  %``K- p scattering length from scattering experiments,''
  Phys.\ Rev.\  C {\bf 74} (2006) 055201
  [arXiv:hep-ph/0606108].
  %%CITATION = PHRVA,C74,055201;%%



\bibitem{KN}
N.~Kaiser, P.~B.~Siegel and W.~Weise,
Nucl.\ Phys.\ A {\bf 594} (1995) 325 
[arXiv:nucl-th/9505043];
%%CITATION = NUCL-TH 9505043;%%

E.~Oset and A.~Ramos,
%``Non perturbative chiral approach to s-wave anti-K N interactions,''
Nucl.\ Phys.\ A {\bf 635} (1998) 99
[arXiv:nucl-th/9711022];
%%CITATION = NUCL-TH 9711022;%%

J.~A.~Oller and U.-G.~Mei{\ss}ner,
Phys.\ Lett.\ B {\bf 500} (2001) 263
[arXiv:hep-ph/0011146].
%%CITATION = HEP-PH 0011146;%%

M.~F.~M.~Lutz and E.~E.~Kolomeitsev,
%``Relativistic chiral SU(3) symmetry, large N(c) sum rules and meson  baryon
%scattering,''
Nucl.\ Phys.\  A {\bf 700} (2002) 193
[arXiv:nucl-th/0105042].

%\cite{Gasser:2007zt}
\bibitem{Gasser:2007zt}
  J.~Gasser, V.~E.~Lyubovitskij and A.~Rusetsky,
  %``Hadronic atoms in QCD + QED,''
  Phys.\ Rept.\  {\bf 456} (2008) 167
  [arXiv:0711.3522 [hep-ph]].
  %%CITATION = PRPLC,456,167;%%

%\cite{Beer:2005qi}
\bibitem{Beer:2005qi}
  G.~Beer {\it et al.}  [DEAR Collaboration],
  %``Measurement of the kaonic hydrogen X-ray spectrum,''
  Phys.\ Rev.\ Lett.\  {\bf 94} (2005) 212302;
  %%CITATION = PRLTA,94,212302;%%

  C.~Curceanu-Petrascu {\it et al.},
  %``Precision Measurements Of Kaonic Atoms At Dafne And Future Perspectives,''
  Eur.\ Phys.\ J.\  A {\bf 31} (2007) 537.
  %%CITATION = EPHJA,A31,537;%%


%\cite{Ito:1998yi}
\bibitem{Ito:1998yi}
  T.~M.~Ito {\it et al.},
  %``Observation of kaonic hydrogen atom x rays,''
  Phys.\ Rev.\  C {\bf 58} (1998) 2366.
  %%CITATION = PHRVA,C58,2366;%%

%\cite{Luscher:1986pf}
\bibitem{Luscher:1986pf}
  M.~L\"uscher,
  %``Volume Dependence of the Energy Spectrum in Massive Quantum Field Theories.
  %2. Scattering States,''
  Commun.\ Math.\ Phys.\  {\bf 105} (1986) 153.
  %%CITATION = CMPHA,105,153;%%

%\cite{Luscher:1990ux}
\bibitem{Luscher:1990ux}
  M.~L\"uscher,
  %``Two particle states on a torus and their relation to the scattering
  %matrix,''
  Nucl.\ Phys.\  B {\bf 354} (1991) 531.
  %%CITATION = NUPHA,B354,531;%%


\bibitem{Beane:2007xs}
  G.~W.~Meng, C.~Miao, X.~N.~Du and C.~Liu,
  %``Lattice study on kaon nucleon scattering length in the I = 1 channel,''
  Int.\ J.\ Mod.\ Phys.\  A {\bf 19} (2004) 4401
  [arXiv:hep-lat/0309048];
  %%CITATION = IMPAE,A19,4401;%%

 S.~R.~Beane, P.~F.~Bedaque, T.~C.~Luu, K.~Orginos, E.~Pallante, A.~Parreno 
 and M.~J.~Savage,
  %``pi K scattering in full QCD with domain-wall valence quarks,''
  Phys.\ Rev.\  D {\bf 74} (2006) 114503
  [arXiv:hep-lat/0607036];
  %%CITATION = PHRVA,D74,114503;%%

  S.~R.~Beane, T.~C.~Luu, K.~Orginos, A.~Parreno, M.~J.~Savage, A.~Torok and A.~Walker-Loud,
  %``Precise Determination of the I=2 pipi Scattering Length from Mixed-Action
  %Lattice QCD,''
  Phys.\ Rev.\  D {\bf 77} (2008) 014505
  [arXiv:0706.3026 [hep-lat]].
  %%CITATION = PHRVA,D77,014505;%%

\bibitem{Aoki:2005uf}
  S.~Aoki {\it et al.}  [CP-PACS Collaboration],
  %``I = 2 pion scattering length from two-pion wave functions,''
  Phys.\ Rev.\  D {\bf 71} (2005) 094504
  [arXiv:hep-lat/0503025].
  %%CITATION = PHRVA,D71,094504;%%



\bibitem{He}
  C.~Liu, X.~Feng and S.~He,
  %``Two particle states in a box and the S-matrix in multi-channel
  %scattering,''
  Int.\ J.\ Mod.\ Phys.\  A {\bf 21} (2006) 847
  [arXiv:hep-lat/0508022].
  %%CITATION = IMPAE,A21,847;%%


\bibitem{cuspwe}
  G.~Colangelo, J.~Gasser, B.~Kubis and A.~Rusetsky,
  %``Cusps in K --> 3pi decays,''
  Phys.\ Lett.\  B {\bf 638} (2006) 187
  [arXiv:hep-ph/0604084];
  %%CITATION = PHLTA,B638,187;%%  

  M.~Bissegger, A.~Fuhrer, J.~Gasser, B.~Kubis and A.~Rusetsky,
  %``Cusps in K_L --> 3 pi decays,''
  Phys.\ Lett.\  B {\bf 659} (2008) 576
  [arXiv:0710.4456 [hep-ph]].
  %%CITATION = PHLTA,B659,576;%%



\bibitem{Lage}
  V.~Bernard, M.~Lage, U.-G.~Mei{\ss}ner and A.~Rusetsky,
  %``Resonance properties from the finite-volume energy spectrum,''
  JHEP {\bf 0808} (2008) 024
  [arXiv:0806.4495 [hep-lat]].
  %%CITATION = JHEPA,0808,024;%%




\end{thebibliography}
\end{document}